\documentstyle[eqsecnum,aps]{revtex}

\begin{document}
\title{Numerical Verification of Percival's Conjecture
in a Quantum Billiard.}
\author{Gabriel Carlo and Eduardo Vergini}
\address{Departamento de F\'\i sica, Comisi\'on Nacional de Energ\'\i a
At\'omica,\\
Avenida del Libertador 8250, 1429, Buenos Aires, Argentina.}

\author {Alejandro J. Fendrik}
\address{Departamento de F\'{\i}sica, Facultad de Ciencias Exactas y
Naturales,\\
Universidad de Buenos Aires, Ciudad Universitaria, 1428, Buenos Aires, \\
Argentina.}
\date{\today}
\maketitle

\begin{abstract}
In order to verify Percival's conjecture [J. Phys. B 6,L229 (1973)] we 
study a planar billiard in its classical and quantum versions. We provide an evaluation of the nearest-neighbor level-spacing distribution for the 
Cassini oval billiard, taking into account relations with classical results. 
The statistical behavior of integrable and ergodic systems has been 
extensively confirmed numerically, but that is not the case for the 
transition between these two extremes. Our system's classical dynamics 
undergoes a transition from integrability to chaos by varying a shape 
parameter. This feature allows us to investigate the spectral 
fluctuations, comparing numerical results with semiclassical predictions 
founded on Percival's conjecture. We obtain good $global$ agreement with 
those predictions, in clear contrast with similar comparisons for other 
systems found in the literature. The structure of some eigenfunctions, 
displayed in the quantum Poincar\'e section, provides a clear explanation 
of the conjecture.

\end{abstract}



\newpage

\section{INTRODUCTION}

\label{sec:int}
In 1973 Percival conjectured that in the semiclassical limit, the 
spectrum of a generic dynamical system consists of two parts with 
strongly contrasting properties: a regular and an irregular part 
\cite{per}. At the classical level such a system exhibits a mixed 
dynamics: regular regions dominated by tori and chaotic regions with 
mixing behavior coexist in the phase space.

In order to characterize a semiclassical spectrum it is advantageous 
to consider a sensitive fluctuation measure. The probability 
distribution $p(s)$ of the spacing $s$ between succesive levels is 
of particular interest because it contains information of the spectrum 
on its finest scale. In the special case of multidimensional integrable
systems, Berry and Tabor \cite{ber1} showed that the levels are
uncorrelated and $p(s)$ is governed by a Poisson distribution. The
other special case corresponds to mixing systems where almost all 
orbits explore densely and chaoticaly the energy surface. In this case, 
Bohigas {\it et al.} \cite{boh} conjectured that the fluctuation 
properties of these spectra can be modeled by the ensemble of random
real symmetric matrices [the Gaussian Orthogonal Ensemble (GOE)] \cite{por}. 
For such systems, where the energy levels display repulsion, $p(s)$ is 
closely approximated by the Wigner distribution; however, we have used the
exact distribution \cite{goe} because the differences are
meaningful, as we shall see below.

In the generic case, Berry and Robnik \cite{ber2}, based on Percival's
conjecture, considered independent sequences of levels 
associated with each connected regular or irregular classical 
phase-space region. When only one chaotic region predominates (the 
situation considered in this article), they provided an expression for 
the distribution $p(s)$ in terms of the classical fraction $\rho^{cl}$ 
of regular regions [referred to as the Berry-Robnik distribution (BRD)].

Although the special cases have been extensively confirmed numerically
\cite{mac,ber3,cassati,boh,boh2,rob}, with some well understood 
exceptions \cite{bal,seb}, for generic systems numerical 
calculations give rise to contradictory conclusions. Recently, there 
has been a number of numerical works \cite{win,hon,pro} showing that 
the Brody distribution (BD) \cite{bro} gives quite 
a satisfactory fit {\it globally}. The Brody distribution is a 
one-parameter family of distributions that interpolates between 
Poisson and Wigner in a simple way; however, it has no semiclassical 
meaning. On the other hand, in 1994 Prosen and Robnik \cite{pro2} 
confirmed numerically semiclassical predictions (the BRD) working on an 
abstract dynamical system : the standard map on a torus. To agree with 
this theory, they needed to compute extremely high excited states 
(around the 30 $10^6$). However, at not so excited states they found 
good global agreement with the Brody distribution both for the 
standard map and for the Limacon-Robnik billiard. In 1995 Prosen 
arrived at the same conclusion working on a two-dimensional 
semiseparable oscillator \cite{pro3}.

The Brody-like behavior at small spacing is understood in terms 
of tunneling between classically separated regions of phase space; 
however, the global agreement with the Brody distribution has no theoretical
foundations. On the other hand, the very slow trasition to semiclassical
predictions can be explained by the presence of partial barriers in
the chaotic regions because the corresponding statistics is not a GOE
for finite $\hbar$ \cite{boh2,boh3}.

The goal of the present article is to verify that the classical support
of eigenfunctions can be clearly identified as regular or chaotic and
this classification is only affected by tunneling between classically
separated regions of phase space (this effect decays exponentially when
$\hbar$ decreases, as it was pointed out in ref. \cite{ber2}).
We compute the spectral fluctuations of a 
one-parameter family of planar billiards : the Cassini ovals 
\cite{cas}. The classical dynamics in this billiard is mixed, 
going from integrability to chaos by varying a single parameter.
We have chosen this parameter such that the classical dynamics 
does not show partial barriers inmerse in the chaotic sea to study 
the Percival's conjecture through the accuracy of the BRD. Moreover, we study
qualitatively the eigenfunctions in phase space to provide additional
support to the results. 

Our work is organized in the following way. In Sec.\ \ref{sec:clas} 
we introduce the classical system. Section \ \ref{sec:quan} is devoted 
to the description of the quantum system (its energy spectrum and the 
corresponding eigenfunctions). In Sec.\ \ref{sec:num} we study the 
resulting energy level statistic. Finally, Sec.\ \ref{sec:con} is 
devoted to conclusions.

\section{THE CASSINI OVAL BILLIARD}

\label{sec:clas}
Our billiard consists of a free-moving point particle inside
a two-dimensional box that bounces off the boundary elastically. 
The boundary of our billiard system is given by a 
fourth-order curve, the Cassini oval:
\[
r_1 r_2=a^2 \;,
\]
where $r_1$ and $r_2$ are distances from two foci located at
$x= \pm c$ and $y=0$. In cartesian coordinates it can be cast into the 
form:
\begin{equation}
(x^2+y^2)^2-2c^2(x^2-y^2)=a^4-c^4 \;. \label{casoval}
\end{equation}
We have two characteristic lengths. However, the shape of the boundary 
is defined by the ratio $d=a/c$ (from now on, the shape parameter), 
which determines the following boundary types: $\sqrt{2}<d$, the boundary 
is an oval; $1<d<\sqrt{2}$, the boundary is an oval with a neck;
$d<1$, the boundary becomes unlinked (two ovals separate).

In the present work we investigate $d>1$ values. Decreasing the shape 
parameter, the classical behavior goes from the regular motion  
(when $d \rightarrow \infty$, the boundary is a circle) to the 
chaotic one (when $d \rightarrow 1$). Using the reflection symmetries 
of the boundary, we consider the motion in the region $x>0,y>0$ 
(desymmetrized billiard). That is, we study the quarter billiard defined 
by the boundary (eq. \ref{casoval}) for $x \geq 0,y \geq 0$ and 
the coordinate axes $x$ and $y$. We study the Cassini oval billiard for 
two values of $d$. One of them is $d=\sqrt{2}$ (the value of the parameter 
for which the neck begins to appear; see Fig.~\ref{shapes}(a)) whose 
form mimics the Bunimovich stadium billiard \cite{buni}.Figure 
\ref{maps}(a) shows the Poincar\'e surface at the boundary using 
Birkhoff coordinates. The coordinate $q$ is related to the arclength 
coordinate at the boundary where the bounce takes place by 
$q=\mbox{(arclength)}/\mbox{(perimeter)}$; and 
$p~=~{\bf p} \cdot {\bf \hat t} / |{\bf p}|$
is the fraction of tangential momentum at this point (${\bf \hat t}$ 
being the tangent unit vector to the boundary). Exploiting the time-reversal 
symmetry we show only the $p \geq 0$, $0 \leq q \leq 1/4$ 
region. The classical phase space has a bouncing-ball regular region 
dominated by invariant curves and a chaotic region with unstable short 
periodic orbits equivalent to those appearing in the Bunimovich stadium 
billiard. A resonance of winding number $6$ defines the last great 
regular region before chaos begin to appear (an eigenfunction existing 
on the chain of islands defined by the resonance is shown in Fig. 
\ref{wavefs} (a)). We have found two very small stable regions 
corresponding to a stable bifurcation of the unstable bow tie periodic 
orbit (two small dark dots can be observed in the chaotic region). 
(Figure \ref{wavefs} (b) shows an eigenfunction existing in that region 
of the phase space). We have not detected any other regular region 
embedded in the chaotic sea. 

The other shape we have studied ($d=2$) is closer to an ellipse 
(see Fig.~\ref{shapes}(b)). In this case the bouncing-ball region
of the phase space is greater than before as it can be seen in
Fig.~\ref{maps}(b). Moreover, a regular region appears as a thin band 
for $p \stackrel {\sim}{>} 0.9$ values, dominated by whispering 
gallery trajectories. Phase space is very mixed and many stable 
islands of very different sizes are interspersed with the chaotic 
trajectories.

By selecting two regions corresponding to chaotic motion in the phase space, 
we have calculated diffusion times between them. The results for  
$d=\sqrt{2}$ are independent of the chosen regions. In the other
case ($d=2$) this time is strongly dependent on them, and we 
have obtained diffussion times one order of magnitude greater than those of
$d=\sqrt{2}$. This fact is related to partial barriers such as those 
shown in Fig.~\ref{barriers}(a).

We have determined $\rho^{cl}$, the fraction of the phase space that 
corresponds to regular motion for both the values of the parameter. We 
have found that $\rho^{cl}= 0.172$ when $d=\sqrt{2}$ while 
$\rho^{cl}= 0.394$ for $d=2$.

\section{THE QUANTUM BILLIARD}

\label{sec:quan}
To study the quantum billiard we solve the time-independent 
Schr\"odinger equation for one particle inside a two-dimensional box 
$\cal{D}$ with Dirichlet boundary conditions at the impenetrable walls 
$\partial{\cal{D}}$:

\[
\nabla ^2 \phi(\vec{r}) = -k^2  \phi(\vec{r}),\mbox{ in $\cal{D}$},
\]
\[
\phi = 0 ,  \mbox{ on $\partial{\cal{D}}$} \;,
\]

where
\[
k = \frac {\sqrt{2mE}} { \hbar }
\] 
and $\cal D$ corresponds to the surface of the desymmetrized billiard,
that is the surface bounded by the Cassini 
oval (\ref{casoval}) and the $x>0$ and $y>0$ semiaxis. So, the  
Dirichlet boundary condition implies that only odd-odd solutions 
of the full billiard will be found. We have employed a new technique,
the scaling method \cite{Ver}: this is a very efficient one dimensional 
method developed to compute eigenvalues and eigenfunctions of quite 
general planar billiards (for three dimensional billiards this is 
practically the only available method to obtain high excited 
states \cite{prosen}). The great advantage of the method is that all 
eigenvalues and eigenfunctions in a narrow $k$ interval are computed 
simultaneously with comparable accuracy, thus avoiding time consuming 
searches and the posibility of missing some state.

We have calculated the energy levels from the fundamental state up to 
the 25000th level for $d=\sqrt{2}$ and up to the 10000th for $d=2$. 
Moreover, we took a sequence of 5000 levels between the 62210th and 
67210th for both values of $d$. We obtained the eigenvalues with an 
average precision of $10^{-6}$ of the mean level spacing for chaotic
eigenfunctions. Regular states are only limited by the computer 
(double) precision.

We have studied the eigenfunctions of the billiard in different 
regions of the spectrum. In general, it is possible to identify each
eigenfunction with a classical region. This identification 
of the eigenfunctions is more clearly seen in the stellar 
representation \cite{voros}. In it, the Husimi distribution of the 
normal derivative on the boundary represents the eigenfunctions in
the Poincar\'e section in Birkhoff coordinates. The stellar 
representation of an eigenfunction is compared directly with the 
classical Poincar\'e section. As an example we show some eigenfunctions
for $d=\sqrt{2}$ (we have taken the area of the desymmetrized billiard 
equal to $\pi / 4$). Fig. \ref{wavefs} (a) shows a linear density plot 
of the square of a state existing on the chain of islands defined by a 
resonance of winding number $6$ and Fig. \ref{husimi} (a) shows the 
same state in stellar representation. Figs. \ref{wavefs} (b) and 
\ref{husimi} (b) show a scar \cite{heller} of the bow tie unstable 
periodic orbit. Figs. \ref{wavefs} (c) and \ref{husimi} (c) show a 
delocalized state distributed over all the classically chaotic region 
(chaotic state). Finally, Figs. \ref{wavefs} (d) and \ref{husimi} (d) 
show a bouncing-ball state.

We stress that the states appearing in Figs. \ref{wavefs} (c) and (d) 
are quasi degenerate. The distance between them is a very small 
fraction ($s=0.00018$) of the mean level spacing. It is clear from 
Figures \ref{husimi} (c) and (d) that these states practically do 
not interact because they exist in different regions of phase space. 
On the other hand, this example shows us that it is necessary to 
evaluate the eigenvalues with high accuracy in order to calculate the 
spectral fluctuations of the system (see the next section).

For $d=2$, the qualitative description of the eigenfunctions is 
equivalent to the previous one. However, there is a significant
fraction of localized eigenfunctions existing in the chaotic region.
One of them is shown in Fig.~\ref{barriers}(b) and (c).

\section{THE ENERGY-LEVEL STATISTIC}

\label{sec:num}

In this section we analyze the level spacing distribution of the 
numerical data described previously. The counting function $N(k)$ gives 
us the number of levels with wave number below $k$. Weyl's formula with 
border, angle and curvature corrections \cite{ber3} provides a good 
estimate for the smooth part $<N(k)>$ of the counting function.

In order to verify that no levels had been lost we have compared 
$<N(k)>$ with a smoothed version of $N(k)$. Defining a new sequence by 
$K_n \equiv <N(k_n)>$ ( where $k_n$ belongs to the original sequence of 
wave numbers ) we take into account of the ``unfolding'' procedure by which 
a unit mean spacing is given to the series of levels \cite{pro2}.

Following ref.\ \cite{pro} in the analysis of the data, we use 
the cumulative level spacing distribution $W(s)=\int_0^s p(y) dy$ 
rather than $p(s)$ because it is numerically easier to evaluate. We fit 
numerical curves with the analytical expression for the Brody family of 
cumulative spacing distributions:
\[
W_\beta^B (s)=1-\exp{(-bs^{\beta+1})} \;,
\]
with $b=(\Gamma((\beta +2)/(\beta +1)))^{\beta +1}$ and  the 
theoretical Berry-Robnik distribution:

\[
W_\rho^{BR} (s)=1-{ \exp{(-\rho s)} [\rho Q_{GOE}(s) - 
 (1- \rho) (W_{GOE}(s) -1)]}
\]
with $Q_{GOE}(s) \equiv (1- \rho) \int_s^\infty d\lambda 
\int_\lambda^\infty p_{GOE}(y) dy$, $p_{GOE}(y)$ being the exact 
GOE spacing distribution, and $W_{GOE} = 1+ dQ_{GOE} / 
d((1- \rho)s)$. $Q_{GOE}$ and $W_{GOE}$ are tabulated 
as $\Psi$ and $F$ (taking as mean density $1- \rho$) respectively 
in reference \cite{goe}, for instance. This exact GOE evaluation 
allows us to distinguish deviations of numerical data from theory 
without including the difference between Wigner and exact GOE formulas 
(which are of approximately the same order; this fact can be verified 
in Fig. \ref{udsqrt2b} (a) which shows the difference between the BRD 
using Wigner surmise and the exact GOE results).

Deviations of numerical data from best-fitting curves can be better 
seen with a transformation defined by:

\begin{equation}
U(W)=(1/\pi) \cos ^{-1} \sqrt{(1-W)} 
\end{equation}

which has constant statistical error over all s. Furthermore, if we 
plot $U(W)$ versus $W_{\beta}^B$ we will have an equally spaced 
distribution of points on the abscissa (see Fig.  \ref{udsqrt2b}, 
for example).

We have evaluated $\chi^2=\sum_{i=1}^N (W(s_i)_{num}-W(s_i)_{theo})^2$ 
weighted with $(\delta W(s_i))^2={W(s_i)(1-W(s_i))}/{N}$ so that we 
could find the optimal values of $\beta$ and $\rho^{BR}$ ($\rho^{BR}$ 
is the resulting best fitting value for the fraction $\rho$ of regular 
levels employed as a free parameter in order to find the lowest 
$\chi^2$). Results from fittings can be seen in Table \ref{table1}. 

It is clear that, for $d=\sqrt{2}$, BRD curve fits data much better
than BD curve. Taking the sequence of 25000 first levels we found that
$\chi^2$ for the BRD was approximately five times lower 
than for the BD. Even for the small $s$ range we could verify a better 
agreement between data and BRD formula than for the BD one, although 
deviations due to tunnelling can be clearly seen in Figure 
\ref{udsqrt2b} (a). For large values of $s$, differences between data 
and BRD could evidence discrepancies with respect to GOE behaviour. 
Taking the sequence of 5000 levels between number 62210 and 67210 we 
got a $\chi^2$ for BRD that was 25 times lower than for BD. This can 
be checked with Fig. \ref{udsqrt2b} (b), where the region of small $s$ 
still shows some deviations from best fitting BRD but the range 
where this happens has become very narrow. In fact, for 
all values of $s$, the agreement is excellent and we can say that 
BRD is working perfectly well at these not so high energy levels.									

In the case of $d=2$, for the first 10000 levels, a same order 
$\chi^2$ was found for the two distributions although was better 
for the BRD than for the BD. In Fig. \ref{ud2b} (a) we can appreciate 
that BD fits numerical data in an acceptable way for small $s$ values. 
This is due to tunnelling effects that persist in a wider $s$ range 
than for the $d=\sqrt{2}$ case. A more complicated structure of 
classical phase space makes tunnelling processes more significant. 
However, for greater $s$ BD no longer follows numerical data, so there 
is no global agreement with it. As in the previous case, we take 5000 
levels between 62210 and 67210. For these energies, $\chi^2$ for BD is 
twice the value for BRD. In Fig. \ref{ud2b} (b) we can see that 
discrepancies in the small $s$ region have reduced, and, though not 
as clear as in $d=\sqrt{2}$ case, numerical data is well adjusted by 
BRD globally.

Finally, we investigated the behaviour of data while going from low to 
high energies. In order to do so we took first 3000 levels for both 
of the shapes; then three stripes of levels, the first, second and 
third 8000 levels for $d=\sqrt{2}$ and the first, second and third 
4000 levels for $d=2$. Results of fittings can be seen in Table 
\ref{table2}. It is clear from Figures \ref{udsqrt2a} and \ref{ud2a}, 
that there is no transition from Brody to Berry-Robnik regime, 
neither for $d=\sqrt{2}$, nor for $d=2$.

\section{SUMMARY AND CONCLUSIONS}
\label{sec:con}

To verify Percival's conjecture, we have studied the quantum version 
of a billiard, depending on one shape parameter $d$. This system shows 
mixed classical dynamics, going from integrability ($d \rightarrow 
\infty$) to chaos ($d \rightarrow 1$) as the parameter is varied. 

On the one hand, the Husimi distribution of
the eigenfuntions that we have obtained clearly displays the classical 
structure of the phase space. Though mixings among regular 
wavefunctions and irregular ones are expected (based on the 
abscence of degeneracies in this one-parameter system that has been 
desymmetrized) they seem to happen only when the energy difference 
is surprisingly small ( even for levels as low as $N \simeq 
2500$ such as the ones exemplified in Fig. \ref{wavefs} (c) and (d)). So, 
we can say that Percival's conjecture is effectively working. Mixed 
functions are exceptions mainly originated in the states whose Husimi 
distributions localize on the last KAM tori. 

On the other hand, the level spacing statistics was fitted by two 
distributions depending on one parameter: the semiclassical 
Berry-Robnik distribution (BRD), which is founded on Percival's 
conjecture, and the Brody distribution (BD). We have found, in all 
cases we have analyzed that the BRD is the best one. Moreover, as 
expected, fits of the BRD are better for increasing values 
of the wave number $k$; however, discrepancies due to 
tunnelling effects are found for small values of $s$, even for the 
best-fitting case (see Fig. \ref{udsqrt2b} (b) - This subject 
is currently under study at the moment). For $d=\sqrt{2}$, the parameters $\rho$ 
of the BRD corresponding to best fits, do not show significant 
differences from the fraction of the regular region 
of the classical phase space. In the case of $d=2$, the best $\rho$'s, 
are systematically greater than the classical value. 

Our calculation of diffusion times shows that the cassini billiard with 
$d=\sqrt{2}$ has only one chaotic region. There are not partial barriers 
dividing chaotic regions of comparable sizes. So we expect that the 
chaotic component of the statistic is given by a single GOE distribution.
In the case $d=2$ we have determined partial barriers
between a main chaotic region and a little one near the whispering gallery
region. The size of the latter is so little that it enlarges (for low energies)
the regular region of the phase space rather than contributes as an 
independent GOE. This fact can be seen in Fig.~\ref{barriers}(b) and (c), 
where we show an eigenfunction localized in this small classical chaotic 
region but having regular characteristics. Moreover, there are small 
regular islands immerse in the chaotic sea that are more relevant for $d=2$ 
(see Fig.~\ref{maps} ) than for $d=\sqrt{2}$. They produce quantum 
localization in their chaotic neighborhood (for instance, see the 
Poincar\'e surface section Fig.~\ref{maps} (a) and the Husimi 
distribution Fig.~\ref{husimi} (b) ) and consequently the resulting 
$\rho$ of the fits, overestimates the fraction corresponding to 
regular motion in the classical phase space. We stress that, in all 
cases, the BD has significant differences from the numerically calculated 
distribution. So we do not observe the BD to BRD transition 
that was seen in other systems.

\section*{ACKNOWLEDGMENTS}

We would like to thank Marcos Saraceno for useful suggestions on the 
original manuscript. This work was partially supported by CONICET and 
UBACyT.

\newpage

\begin{figure}
\caption{Desymmetrized Cassini oval (upper-right quarter of the curve 
with segments of the coordinate axes $x$ and $y$). a) $d=\sqrt{2}$ with 
Birkhoff coordinates $p$ and $q$. b) $d=2$.}
\label{shapes}
\end{figure}

\begin{figure}
\caption{Poincar\'e surface of section expressed in Birkhoff 
coordinates ($p$ and $q$) for $p \geq 0$ and $0 \leq q \leq 0.25$. 
a) $d=\sqrt{2}$. b) $d=2$.}
\label{maps}
\end{figure}

\begin{figure}
\caption{Wave functions of Cassini oval billiard for $d=\sqrt{2}$ with 
wave numbers $k$ displayed below each one of them. We can see: 
a) regular eigenfunction, b) eigenfunction strongly localized on a scar 
reminiscent from an unstable periodic orbit (bow tie), c) irregular 
wavefunction and d) bouncing-ball wavefunction extremely close in energy 
to the previous one but without exhibiting mixture.}
\label{wavefs}
\end{figure}

\begin{figure}
\caption{Husimi plots corresponding to the wave functions displayed in 
Figure 3. This shows: a) great localization on a classical resonance, 
b) remarkable localization on the classical region where an unstable 
orbit is found to exist, c) almost uniform extension over chaotic 
region and d) strong localization on one torus belonging to the 
interior of the principal regular island situated at the origin.}
\label{husimi}
\end{figure}

\begin{figure}
\caption{Effects of barriers: a) Poincar\'e section for $d=2$ taking 
only one initial condition very close to the whispering gallery region 
($p \stackrel {\sim}{>} 0.9$). We can observe two partial barriers 
limiting flux and defining two small chaotic regions (which are filled 
more densely than the major chaotic part of phase space). b) Husimi 
distribution of an eigenfunction localized over one of these small 
regions. c) Same eigenfunction in configuration space.}
\label{barriers}
\end{figure}

\begin{figure}
\caption{Differences between numerical $U$ and $U^{BR}$ taking the 
best-fitting value $\rho^{BR}$ are displayed with a solid (fluctuating) 
line for $d=\sqrt{2}$. Dotted lines that follow the solid one represent 
the $\delta U$ uncertainty band. a) $N$ between 1 and 25000 and 
b) $N$ between 62210 and 67210. In a), difference 
between $U^{BR}$ calculated with Wigner surmise and with exact 
GOE results is also plotted with a dashed line. We can see that 
deviations of numerical data from this curve are of the same order.}
\label{udsqrt2b}
\end{figure}

\begin{figure}
\caption{Same as in Figure 6 but for $d=2$. In a) $N$ between 
1 and 10000 and b) $N$ between 62210 and 67210 (we use the same line 
patterns as in Figure 5).}
\label{ud2b}
\end{figure}

\begin{figure}
\caption{Same as in Figure 6, but for $d=\sqrt{2}$. The levels used for 
plottings are: a) 1-3000, b) 1-8000, c) 8001-16000 and 
d) 16001-24000.}
\label{udsqrt2a}
\end{figure}

\begin{figure}
\caption{Same as in Figure 7, but for $d=2$ . Levels used in this case 
are: a) 1-3000, b) 1-4000, c) 3000-7000 and d) 6000-10000.}
\label{ud2a}
\end{figure}

\newpage

\begin{table}
\caption{Results of fittings. Comparison between regular fraction of 
levels $\rho^{BR}$ and integrable part of classical phase space 
$\rho^{cl}$.}
\begin{tabular}{||l|l|l|l|l|l|l||} \hline
$d$ & N & $\beta^B$ & $\chi^2/N$ & $\rho^{BR}$ & $\chi^2/N$ 
& $\rho^{cl}$ \\ \hline
$\sqrt{2}$ & 1 - 25000 & 0.570 & 19.61 & 0.177 & 4.34 & 0.172 \\ 
\hline
$\sqrt{2}$ & 62210 - 67210 & 0.599 & 4.50 & 0.165 & 0.20 & 0.172 
\\ \hline
2 & 1 - 10000 & 0.230 & 5.63 & 0.433 & 4.79 & 0.394 \\ \hline
2 & 62210 - 67210 & 0.226 & 2.27 & 0.437 & 0.98 & 0.394 \\ \hline
\end{tabular}
\label{table1}
\end{table}

\begin{table}
\caption{Results of fittings. Three stripes of increasing energy 
are displayed for the two values of shape parameter.}
\begin{tabular}{||l|l|l|l|l|l||} \hline
$d$ & N & $\beta^B$ & $\chi^2/N$ & $\rho^{BR}$ & $\chi^2/N$ \\ \hline
$\sqrt{2}$ & 1 - 8000 & 0.59 & 10.42 & 0.17 & 3.15  \\ \hline
$\sqrt{2}$ & 8001 - 16000 & 0.57 & 5.79 & 0.17 & 1.33  \\ \hline
$\sqrt{2}$ & 16001 - 24000 & 0.54 & 4.65 & 0.19 & 1.32  \\ \hline
2 & 1 - 4000 & 0.24 & 3.00 & 0.42 & 2.88  \\ \hline
2 & 3000 - 7000 & 0.22 & 2.68 & 0.44 & 1.02  \\ \hline
2 & 6000 - 10000 & 0.22 & 1.61 & 0.45 & 1.97  \\ \hline
\end{tabular}
\label{table2}
\end{table}


\newpage

\begin{references}

\bibitem{per} I. C. Percival, J. Phys. B {\bf 6}, L229 (1973).

\bibitem{ber1} M. V. Berry and M. Tabor, Proc. R. Soc. {\bf A 356},
375 (1977).

\bibitem{boh} O. Bohigas, M-J Giannoni and C. Schimit, Phys. Rev. 
Lett. {\bf 52}, 1 (1984).

\bibitem{por} C. E. Porter, {\it Statistical Theories of Spectra:
Fluctuations} (New York: Academic, 1965).

\bibitem{goe} M. Gaudin, Nuclear Physics {\bf 25}, 447 (1961).

\bibitem{ber2} M. V. Berry and M. Robnik, J. Phys. A {\bf 17}, 2413 
(1984).

\bibitem{mac} S. W. McDonald and A. N. Kaufman, Phys. Rev. Lett.
{\bf 42}, 1189 (1979).

\bibitem{ber3} M. V. Berry, Ann. Phys., $NY$ {\bf 131}, 163 (1981).

\bibitem{cassati} G. Casati, F. Valz Gris, and I. Guarneri, Nuovo 
Cimento Lett. {\bf 28}, 279 (1980).

\bibitem{boh2} O. Bohigas, S. Tomsovic and D. Ullmo, Phys. Rev. Lett.
{\bf 65}, 5 (1990).

\bibitem{rob} M. Robnik, J. Phys. A {\bf 25}, 1399 and 3593 (1992).

\bibitem{bal} N. L. Balazs and A. Voros, Phys. Rep. {\bf 143}, 109 
(1986).

\bibitem{boh3}  O. Bohigas, M-J Giannoni and C. Schimit, {\it Quantum 
Chaos and Statistical Nuclear Physics, Lecture Notes in Physics}, 
T.H. Seligman and H. Nishioka eds., vol. 263 (Berlin: 
Springer, 1986), 18. O. Bohigas, S. Tomsovic and D. Ullmo, 
Phys. Rep. {\bf 223, }, 44 (1993).

\bibitem{seb} P. Seba, Phys. Rev. Lett. {\bf 64}, 1855 (1990).

\bibitem{win} D. Wintgen and H. Friedrich, Phys. Rev. A {\bf 35},
1464 (1987).

\bibitem{hon} A. Honig and d. Wintgen, Phys. Rev. A {\bf 39}, 5642 
(1989).

\bibitem{pro} T. Prosen and M. Robnik, J. Phys. A {\bf 26}, 2371 
(1993).

\bibitem{bro} T. A. Brody, Lett. Nuovo Cimento {\bf 7}, 482 (1973).

\bibitem{pro2} T. Prosen and M. Robnik, J. Phys. A {\bf 27}, 8059 
(1994).

\bibitem{pro3} T. Prosen, J. Phys. A {\bf 28}, L349 (1995).

\bibitem{cas} A. J. Fendrik, J. L. Vega, C. O. Dorso and M. Bernath, 
Nonlinear Phenomena in Fluids, Solids and other Complex Systems, 
P. Cordero, B. Nachtergaele (Eds.), Elsevier Science Publishers B.V.,
 447 (1991).

\bibitem{buni} L. A. Bunimovich, Commun. Math. Phys. 
{\bf 65}, 295 (1979).

\bibitem {Ver} E. Vergini and M. Saraceno, Phys. Rev. E {\bf 52}, 
2204 (1995).

\bibitem {prosen} T. Prosen, Preprint Chao-dyn/9611015, (1996).

\bibitem {voros} J.M. Tualle and A. Voros, Chaos, Solitons and 
Fractals {\bf 5}, 1085 (1995).

\bibitem {heller} E. J. Heller, Phys. Rev. Lett. {\bf 53}, 1515 
(1984).



\end{references}
\end{document}